\documentstyle[twocolumn,aps]{revtex}
  \input epsf
\begin{document}
\draft
\twocolumn[\hsize\textwidth\columnwidth\hsize\csname@twocolumnfalse\endcsname

\title{Variety and Volatility in Financial Markets}
\author{Fabrizio Lillo and Rosario N. Mantegna} 
\address{ Istituto
Nazionale per la Fisica della Materia, Unit\`a di Palermo\\ and\\
Dipartimento di Fisica e Tecnologie Relative, Universit\`a  di
Palermo, Viale delle Scienze, I-90128, Palermo, Italia}
\date{\today} 
\maketitle %\receipt{}

%----------------------------------------------------------------------
\begin{abstract} We study the price dynamics of stocks traded in a
financial  market by considering the statistical properties both of
a single time series and of an ensemble of stocks traded
simultaneously.  We use the $n$ stocks traded in the New York Stock
Exchange to  form a statistical ensemble of daily stock returns. 
For each trading day of our database, we study the
ensemble return distribution. We find that a typical ensemble
return distribution exists in most of the trading days with the
exception of crash and rally days and of the days subsequent to
these extreme events. We analyze each ensemble return distribution
by extracting its first  two central moments. We observe that these
moments are fluctuating  in time and are stochastic processes
themselves. We characterize  the statistical properties of ensemble
return distribution central  moments by investigating  their
probability density functions and temporal correlation properties.
In general, time-averaged and portfolio-averaged price returns have 
different statistical properties. We infer from these differences 
information about the relative strength of correlation between stocks 
and between different trading days. Lastly, we compare our empirical 
results with those predicted by the single-index model and we conclude that 
this simple model is unable to explain the statistical properties of the 
second moment of the ensemble return distribution. 
\end{abstract}
%----------------------------------------------------------------------
 \vskip2pc]

%\newpage
\narrowtext
\section{INTRODUCTION}

In recent years physicists started to interact with economists to 
concur to the modeling of financial markets as
model complex systems \cite{Anderson88}. This triggered the interest
of a group of physicists into the analysis and modeling of price 
dynamics in financial markets performed by using paradigms
and tools of statistical and theoretical physics \cite{Proc}. 
One target of these researches is to implement a stochastic 
model of price dynamics in 
financial markets which reproduces the statistical properties
observed in the time evolution of stock prices.
In the last few years physicists interested in financial analysis 
have performed several empirical researches investigating the 
statistical properties of stock price and volatility time series 
of a single stock (or of an index) at different temporal 
horizons \cite{Ms99,Bp00}. Such a kind of analysis does not take
into account any interaction of the considered financial stock 
with other stocks traded simultaneously in the same market.
It is known that the synchronous price returns time series of 
different stocks are pair correlated \cite{Elton,Campbell} and several
researches has been performed also by physicists in order to 
extract information from the correlation properties 
\cite{Mantegna99,Bouchaud99,Stanley99}. A precise
characterization of collective movements in a financial market 
is of key importance in understanding the market dynamics and in 
controlling the risk associated to a portfolio of stocks. The present study 
contributes to the understanding of collective behavior of a portfolio of 
stocks in normal and extreme days of market activity.  

Specifically, we address the question: Is the complexity 
of a financial market essentially limited to the
statistical behavior of each financial time series or 
rather a complexity of the overall market exists?
To answer this question, we present the results of an empirical 
analysis performed adopting the following point of 
view. We investigate the price returns of an 
ensemble of $n$ stocks simultaneously traded in a financial 
market at a given day. With this approach we
quantify what we call the {\it variety} of a financial market 
at a given trading day \cite{Lillo99}. The variety provides 
statistical information about the amount of
different behavior observed in stock return in a given ensemble 
of stocks at a given trading time horizon (in the present 
case, one trading day). We observe that the distribution of variety 
is sensitive to the composition of the portfolio investigated 
(especially to the capitalization of the considered stocks).

The return distribution shows a typical shape for most of the trading days. 
However, the 
typical behavior is not observed during crash and rally days.
The shape and parameters characterizing the ensemble return distribution
are relatively stable during normal phases of the market activity 
while become time dependent in the periods subsequent to crashes.
The variety is characterized by a long-range correlated memory
showing that no typical time scale can be expected after a rally or
a crash for the expected relaxation to a ``normal" market phase.
Moreover a simple model such as the
single-index model is not able to reproduce the statistical properties 
empirically observed.

The paper is organized as follow. In Section II we illustrate  
our database and the ensemble of stocks considered. 
Sect. III is devoted to the investigation of the statistical 
properties of the time evolution of each single stock.
In Section IV, we 
discuss the statistical properties of ensemble return distribution.
Specifically we consider the behavior of the central lowest moments, 
their distribution and correlation, a comparison of time and
portfolio average, and the role of the size and homogeneity of
the investigated portfolio.  
In Section V we compare the statistical properties observed
in a real financial market with the prediction of the
single-index model. In 
Section VI we present a discussion of the obtained results.

\section{DATABASE AND INVESTIGATED VARIABLES}

The investigated market is the New York Stock Exchange (NYSE) 
during the 12-year period from January 1987 to December 1998 
which corresponds to 3032 trading days.  
We consider the ensemble of all stocks traded in the NYSE. 
The number of stocks traded in the NYSE is increasing
in the investigated period and it ranges from $1128$ at 
the beginning of 1987 to $2788$ at the end of 1998.
The total number of data records exceeds 
$6$ millions. 

The variable investigated in our analysis is the daily price 
return, which is defined as
\begin{equation}
R_i(t)\equiv\frac{Y_i(t+1)-Y_i(t)}{Y_i(t)},
\end{equation}  
where $Y_i(t)$ is the closure price of $i-$th stock at 
day $t$ ($t=1,2,..$). For each trading day $t$, 
we consider $n$ returns, where $n$ is depending on the 
total number of stocks traded in the NYSE at the selected
day $t$. In our study we use a ``market time". With this choice, 
we consider only the trading days and we remove the weekends 
and the holidays from the calendar time.

A database of more than 6 millions records unavoidably 
contains some errors. A direct control of a so large 
database is not realistic. For this reason,  to avoid 
spurious results we filter the data by not considering 
price returns which are in absolute values greater than 
$50\%$. 

The companies traded in the NYSE are quite different the one from
the other. Differences
among the companies are observed both with respect to the sector
of their economic interests   and with respect to their size. One
measure of the size of a  company is its capitalization.  
The capitalization of a stock is the stock price
times the number of outstanding shares. In this study, 
we discuss the role of the different capitalization in the  price
dynamics.

\section{SINGLE stock PROPERTIES}

The distribution of returns with different time 
horizons of a single stock or index has been studied 
by several authors \cite{Proc,Ms99,Bp00,Elton,Campbell}.

The stocks traded in a financial market have 
different capitalization. An important point is whether 
the differences in capitalization are reflected in the 
statistical properties of the price returns of the stocks. 
To answer this question we investigate the distribution of 
daily returns of $2188$ stocks traded in the NYSE 
at an arbitrarily chosen day that we select as June 10th, 1996.

\begin{figure}[t]
\epsfxsize=3in
\epsfbox{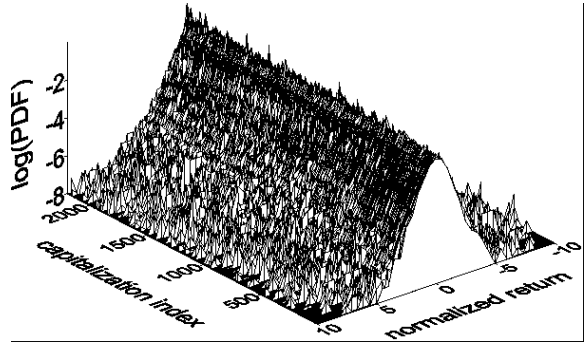}
\vspace{0.3cm}
\caption{Surface plot of the logarithm of the probability density function
of normalized daily returns $(R_i(t)-\mu_i)/\sigma_i$ of all the stocks traded 
in the NYSE. The stocks are sorted according to their capitalization at June 
10th, 1996.}
\label{fig1}
\end{figure}

We compare the statistical properties of daily price return
distribution of each stock as a function of its capitalization. 
We order the $2188$ stocks in decreasing order
according to their capitalization at June 10th, 1996.
Our ordering procedure gives 
to the most capitalized stock (the General Electric Co., GE) 
the rank $i=1$, to the second one (the Coca Cola 
Company) the rank $i=2$, and so on. An analysis of the 
return probability density function (pdf) for the 
$2188$ stocks shows that the distributions are different. 
This is due in general to: (i) different scale and
(ii) different shape of the return pdfs. In order 
to eliminate one source of difference we analyze 
the pdf of the normalized
returns $(R_i(t)-\mu_i)/\sigma_i$ $(i=1,2,...,2188)$,
where $\mu_i$ and $\sigma_i$ are the 
first two central moments of the time series $R_i(t)$ defined as
\begin{eqnarray}
&&\mu_i=\frac{1}{T_i}\sum_{t=1}^{T_i} R_i(t), \\
&&\sigma_i= \sqrt{\frac{1}{T_i}\left(\sum_{t=1}^{T_i} (R_i(t)-\mu_i)^2\right)},
\end{eqnarray} 
where $T_i$ is the number of trading days of the stock $i$
during the investigated period. 
The quantity $\mu_i$ gives 
a measure of the overall performance of stock $i$ 
in the period. The standard deviation $\sigma_i$ is 
called {\it historical volatility} in the financial literature
and quantifies the risk associated with the $i$-th stock. 
This quantity is of primary importance in risk management 
and in option pricing. 

The pdf of normalized daily returns 
of all the stocks
ordered by capitalization is shown in Fig. 1. 
The central part of the distribution of the most
capitalized stocks has 
\begin{figure}[t]
\epsfxsize=3in
\epsfbox{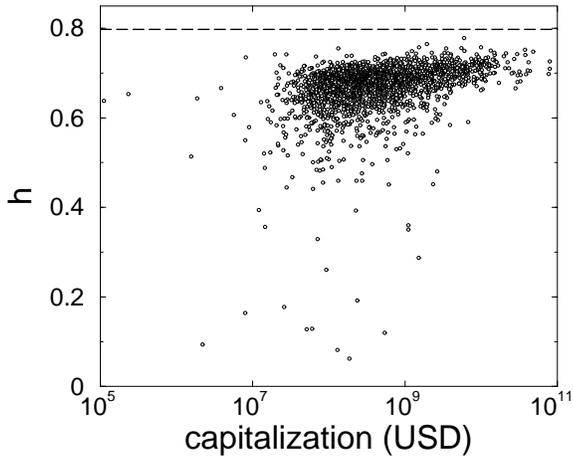}
\caption{Each circle represents the $h$ parameter defined in Eq. (4) 
of the daily return distribution of a stock as a function of its capitalization.
The dashed line is the value $\sqrt{2/\pi} \simeq 0.80$ which is the lower
bound for $h_G$ expected for a Gaussian distribution 
of daily return. Values of $h$ smaller than $h_G$ indicate a leptokurtic 
distribution of returns. The parameter $h$ slowly increases by increasing 
the capitalization.}
\label{fig2}
 \end{figure}
a bell-shaped profile.  
Moving towards less capitalized stocks the central 
part of the distribution becomes more peaked and the 
tails of the distribution become fatter. The pdf of 
the less capitalized stocks is therefore more leptokurtic 
than the pdf of the more capitalized ones. 
 
The typical estimation of the degree of leptokurtosis 
of a pdf is done by considering its kurtosis. 
The evaluation of the kurtosis of the pdf is in general 
difficult for small set of data because the fourth moment 
and all the moments higher than the second are extremely 
sensible to the highest absolute returns. This implies that 
the kurtosis calculated from a relatively small set of records
is dominated by the highest absolute returns rather than by 
the shape of the pdf and therefore it is not a good statistical
estimation. To avoid this problem, 
we quantify the distance between the empirically calculated pdf
of daily returns of $i-$th stock and the Gaussian 
distribution by considering the quantity
\begin{equation}
h \equiv \frac{<|x|>}{\sqrt{<x^2>-<x>^2}}.
\end{equation} 
The quantity $h$ is nondimensional and depends on the first two
moments. For the Gaussian distribution 
\begin{equation}
P_G(x)=\frac{1}{\sqrt{2\pi\sigma_G ^2}} \exp \left(-\frac{(x-\mu_G)^2}
{2\sigma_G ^2}
\right),
\end{equation}
the parameter $h$ is equal to 
\begin{equation}
h_G=\sqrt{\frac{2}{\pi}}\left(\exp(-\frac{\mu_G^2}{2\sigma_G^2})+\sqrt{\frac{\pi}{2}}
\frac{\mu_G}{\sigma_G}Erf\left(\frac{\mu_G}{\sqrt{2}\sigma_G}\right)\right).
\end{equation}
The parameter $h_G$ is a function of the ratio $\mu_G/\sigma_G$ ranging
from the lower bound $\sqrt{2/\pi}$ when $\mu_G/\sigma_G=0$ to infinity.
\begin{figure}[t]
\epsfxsize=3in
\epsfbox{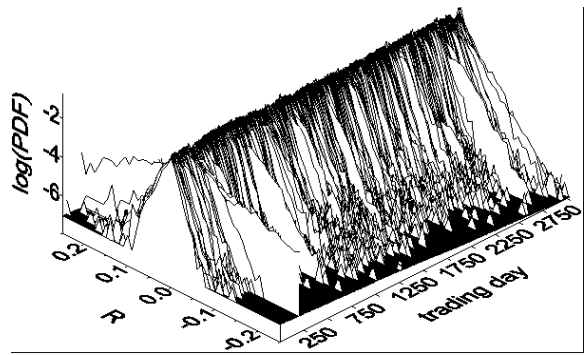}
\vspace{0.3cm}
\caption{Surface plot of the logarithm of the ensemble return distribution
for the 12-year investigated period from January 1987 to December 1998. 
From the Figure is clearly recognizable the 1987 crash (trading day index 
equal to 200) and the high volatility two-year period 1997-1998 (trading
day index from 2500 to 3032).}  
\label{fig3}
 \end{figure}
For a leptokurtic pdf, as for example a Laplace distribution or a
Student's  t-distribution with finite variance, $h$ is always
smaller than $h_G$. The distance of $h$ from $h_G$ is able to
quantify the degree  of leptokurtosis of the considered pdf.      
Figure 2 shows the parameter $h$ for the  stocks traded in the NYSE
as a function of their capitalization.  In the figure, we show also
the lower bound of $h_G$ for comparison.  The empirically
calculated parameter $h$ is systematically smaller than $h_G$. 
The mean value $<h>$ of the overall market is $<h>=0.67$ and its
standard deviation is $\sigma_h=0.06$. Hence this result suggests
that as a first approximation one can assume that the large majority
of stocks are characterized by a roughly similar pdf. However we wish to point
out that this conclusion is only valid as a first approximation because a trend
of $h$ is clearly detected in Fig. 2. Specifically $h$ increases as
the capitalization increases.  Therefore the less capitalized
stocks have a  more leptokurtic daily return pdf than the more
capitalized ones.

The second moment of return distribution has been found
finite in recent research \cite{Akgiray88,Mantegna95,Lux96,Gopikrishnan98}.
In order to verify the convergence of the pdf towards a 
Gaussian pdf at large temporal horizons, we evaluate 
the $h$ parameter for weekly $<h_w>$ and monthly $<h_m>$ return
pdfs. We obtain from our analysis $<h_w>=0.70$ and $<h_m>=0.74$. 
These results show that the values of $h$ moves towards 
$h_G=\sqrt{2/\pi} \simeq 0.80$ when the time horizon of returns 
is increased, supporting the conclusion of finite second moment. 

\section{ensemble return distribution}

In the previous section we focused on statistical properties of time 
evolution of price returns for each single stocks traded in the
NYSE. In this section we perform a synchronous analysis on the return 
of all the stock traded in the NYSE. To this aim we extract the $n$ 
returns of the $n$ stocks for each trading day $t$. 
\begin{figure}[t]
\epsfxsize=3in
\epsfbox{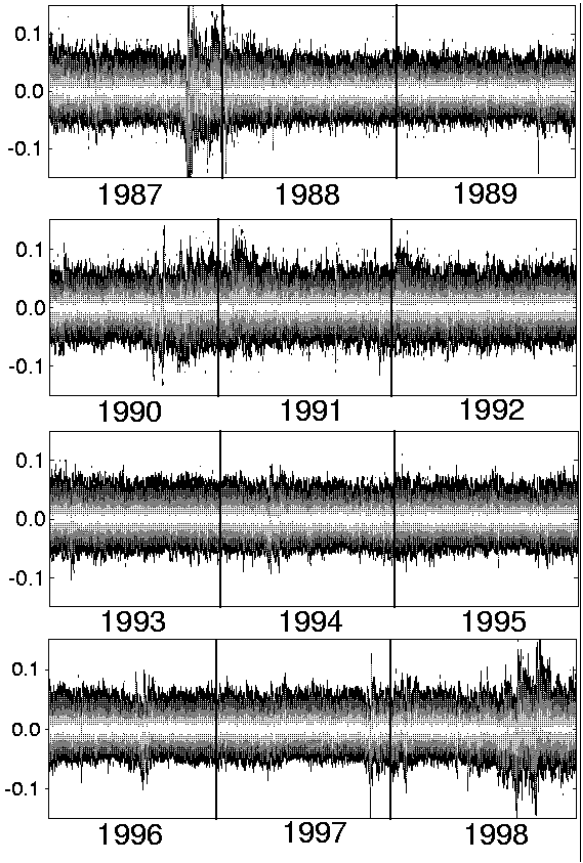}
\vspace{3.6cm}
\caption{Contour plot of the logarithm of the ensemble return distribution
for the 12-year investigated period from January 1987 to December 1998
(same data as in Fig. 3). The contour plot is obtained for equidistant 
intervals of the logarithmic probability density. The brightest area of 
the contour plot corresponds to the most probable value.}
\label{fig4}
 \end{figure}
The distribution of these
returns $P_t(R)$ provides  information about the kind of activity
occurring in the market  at the selected trading day $t$.    

Figure 3 shows the logarithm of the pdf 
as a function of the return and of the trading day.
In this figure we show the interval of daily returns from 
$-25\%$ to $25\%$. The central part of the distribution 
is roughly triangular in a logarithmic scale 
and this shape and its scale are conserved for long time 
periods. Sometimes the shape and scale of the ensemble return 
pdf changes abruptly either in the presence of large average 
positive returns or large average negative returns.
Figure 4 shows the same data of Fig. 3 in a contour plot.
The contour lines describe equiprobability regions. In order to 
point out the properties of the central part of the distribution, 
in Fig. 4 we plot only 
the returns which are less than $15\%$ in absolute value. 
Only a few points of the contour lines fall behind this limit 
during the 1987 and 1998 crises.
In Fig. 4 there are long time periods in which the 
central part of the distribution maintains his shape and the 
equiprobability contour lines are approximately parallel one 
to each other. As an example, one can consider the 
three-year period 1993-1995. 
\begin{figure}[t]
\epsfxsize=3in
\epsfbox{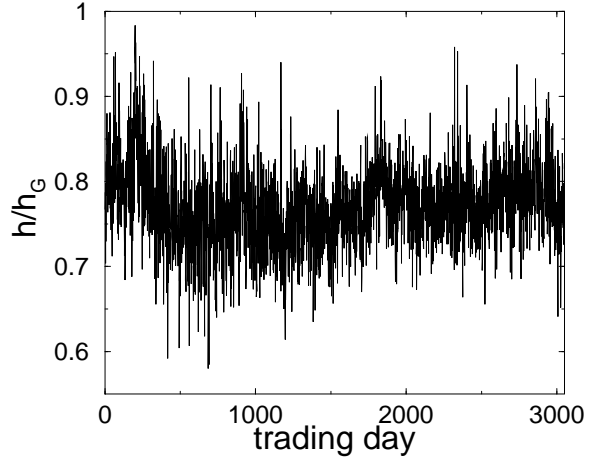}
\caption{Ratio between $h$ parameter defined in Eq. (4) of the ensemble 
return distribution and the value of $h_G$ expected by a Gaussian 
distribution and defined by Eq. (6) for each trading day. 
The ratio $h/h_G$ is systematically smaller than one, indicating 
that the ensemble return distribution is leptokurtic for each trading day.}
\label{fig5}
 \end{figure}
On the other hand there are 
time periods in which the shape of the distribution changes 
drastically. In general these periods corresponds 
to financial turmoil in the market. For example 
a dramatic change of the shape and of the scale of the pdf
is observed in Fig. 4 during and after the 19 Oct. 1987 crash,
at the beginning of 1991 and at the end of 1998.
A systematic analysis of the change of the shape and scale 
of the ensemble return distribution during extreme events of the
market has been discussed elsewhere \cite{epjb00} 

One key aspect of the ensemble return distribution concerns its
shape during the normal periods of activity of the market. Is the
distribution approximately Gaussian or systematic deviation from a
Gaussian shape are quantitatively observed? We already cited  that
a direct inspection of Fig. 3 suggests that the central part of
the empirical return distribution is roughly Laplacian  (triangular
in a logarithmic scale) and not Gaussian. To make this analysis
more quantitative, we show in Fig. 5 the ratio between the
value of $h$ determined for each trading day from the ensemble
return distribution and the quantities $h_G$ calculated by determining
the mean and the standard deviation of $P_t(R)$ and hypothesizing
a Gaussian shape by using Eq. (6). The ratio $h/h_G$
is systematically smaller than one and this implies that the Gaussian
hypothesis for the shape of the distribution is not verified by
the empirical analysis. In other words the Gaussian distribution
is not a good approximation  both for the central part and for the
tails of the distribution and the deviation from the Gaussian
behavior is systematically observed for all the trading days of
the 12 years time period analyzed in our study.

In summary the ensemble return distribution well characterizes the market
activity. It has a typical shape and scale during long periods of
``normal" activity of the 
\begin{figure}[t]
\epsfxsize=3in
\epsfbox{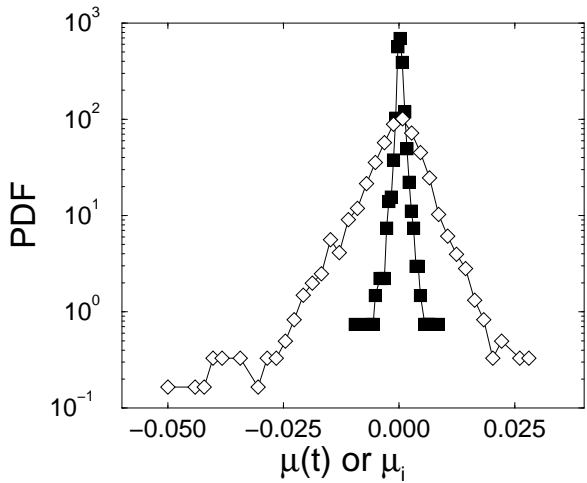}
\vspace{0.5cm}
\caption{Linear-log plot of the probability density function of the
mean $\mu(t)$ of the ensemble return distribution (white diamond) 
and of the mean of the daily return $\mu_i$ of all the stocks traded in the 
NYSE (black square).}
\label{fig6}
 \end{figure}
market characterized by moderately low
average  daily return. During extreme events the shape and scale
are  dramatically changed in a systematic way. Specifically during
crises the ensemble return distribution becomes negatively skewed
whereas during rallies a positive skewness is observed
\cite{epjb00}.  Figure 4 clearly shows that extreme events (such as
for example October 87 crash) triggers an ``aftershock" period, 
in the ensemble return pdf, that can last for a
period of time of several months. 

\subsection{Central moments}

In order to characterize more quantitatively  the ensemble return
distribution at day $t$, we extract the  first two central moments
at each of the $3032$  trading days. Specifically, we consider the
average  and the standard  deviation defined as
\begin{eqnarray}
&&\mu (t)=\frac{1}{n_t}\sum_{i=1}^{n_t} R_i(t), \\
&&\sigma (t)= \sqrt{\frac{1}{n_t}\left(\sum_{i=1}^{n_t}
(R_i(t)-\mu(t))^2\right)},
\end{eqnarray}
where $n_t$ indicates the number of stocks traded at day $t$.

The mean of price returns $\mu(t)$ quantifies  the general trend
of  the market at day $t$.  The standard deviation $\sigma(t)$
gives a measure of the width of  the ensemble return distribution.
We call this quantity {\it variety} of the  ensemble because it
gives a measure of the variety of behavior observed in a
financial market at a given day. A large value of $\sigma(t)$
indicates that different  companies are characterized by rather
different returns at day $t$.  In fact in days of high variety
some companies perform great gains 
\begin{figure}[t]
\epsfxsize=3in
\epsfbox{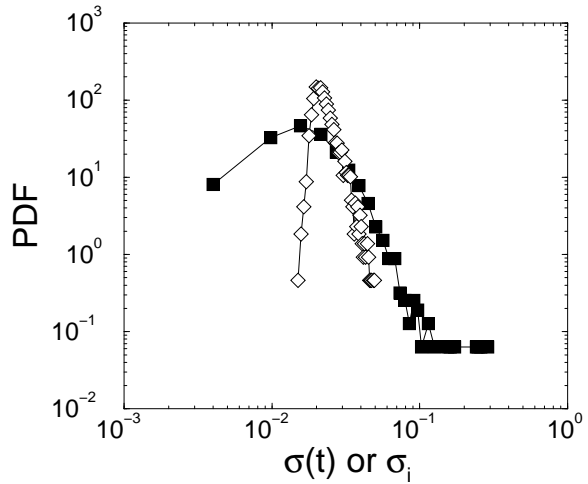}
\vspace{0.5cm}
\caption{Log-log plot of the probability density function of the
variety $\sigma(t)$ , i.e. the  variance of the ensemble return
distribution (white diamond) and of the volatility
$\sigma_i$, i.e. the variance of the daily return, of the all the
stocks  traded in the NYSE (black square).}
\label{fig7}
 \end{figure}
whereas others have great
losses. The mean and the standard
deviation of price returns are not constant and fluctuate in time.
We study the temporal series of $\mu(t)$ and $\sigma(t)$ in order to 
characterize the temporal evolution of the ensemble
return  distribution quantitatively. 
We investigate these fluctuating parameters
by investigating their time correlation properties and their
pdfs.

\subsection{Probability distributions of the central moments}

The empirical pdf of the mean $\mu(t)$
for the 3032 trading days investigated  is shown in Fig. 6. 
The central part of this distribution is non-Gaussian 
and is roughly described by a Laplace 
distribution. 

The mean $\mu(t)$ is proportional to the sum of $n$ random
variables $R_i(t)$ $(i=1,2,...,n)$. The Central Limit Theorem
prescribes  that the sum of $n$ {\it independent} random variables
with finite  variance converges to a Gaussian pdf.  By assuming a
finite value for the volatility of stocks, the observation that 
the pdf of the mean return $\mu(t)$ is non-Gaussian can be
therefore attributed to the presence of correlation between the
stocks.

Figure 7 shows the pdf of the variety
$\sigma(t)$.  The central part of this distribution is
approximated by a lognormal distribution. A deviation from the lognormal
behavior is observed in the tail of higher values of variety.
This deviation is depending on the size of the portfolio and will be discussed
in subsection IV E.

\subsection{Correlations in the central moments}

Another important statistical property of $\mu(t)$ and $\sigma(t)$
concerns their correlation properties \cite{Lillo99}.
For the considered portfolio, we 
calculate the autocorrelation function of a variable $x(t)$
which is defined as  
\begin{equation}
R(\tau)\equiv\frac{<x(t)x(t+\tau)>-<x(t)><x(t+\tau)>}{<x(t)^2>-<x(t)>^2}.
\end{equation} 

In agreement with previous results \cite{Lillo99}, we find that 
the mean $\mu(t)$ is approximately delta correlated, 
whereas the autocorrelation function of $\sigma(t)$ is long-range 
correlated. The empirical autocorrelation function of 
$\sigma(t)$ is well approximated by a power-law function 
$R(\tau)\propto \tau ^{-\delta}$ . By performing a best fit with 
a maximum time lag of $50$ trading days, we determine the exponent 
$\delta=0.230 \pm 0.006$.
This result indicates that the variety $\sigma(t)$ has a long-time memory in 
the market. We recall that the
historical volatility is characterized by long time memory of
the same nature \cite{Dacorogna,Stanley,Serva}.

Another way to investigate the long-range correlation is to determine
the power spectrum of the investigated variable. 
We evaluate the power spectrum of $\sigma(t)$ 
and we perform a best fit of the power spectrum with a functional 
form of the kind
\begin{equation}
S(f)\propto \frac{1}{ f^{\eta}}.
\end{equation}
Our best fit for the power spectrum of $\sigma(t)$ gives for 
the exponent $\eta\approx 1.1$. 
This result confirms that the variety $\sigma(t)$ is
a long-range time correlated random variable.

\subsection{Time and portfolio average}

Figure 6 shows two curves. In fact in Fig. 6 we also show the
pdf of the mean $\mu_i$. The quantity
$\mu_i$ (see Eq. (2)) is the mean return of stock $i$ averaged over
the investigated time interval. The pdf of $\mu_i$ is
non-Gaussian and it is much  more peaked than the pdf of $\mu(t)$.
Hence the statistical behavior observed by investigating a  large
portfolio in a market day is not representative of  the statistical
behavior observed by investigating the time evolution  of single
stocks. 

This comparison can be performed also for the second moment of the
distributions. In Fig. 7 we compare the pdf of the  volatility
$\sigma_i$ and the pdf of the variety $\sigma(t)$. Also in this
case, the statistical properties of $\sigma_i$  and $\sigma(t)$ are
different. Specifically, the pdf of $\sigma(t)$  is more peaked
than the pdf of $\sigma_i$.

In order to understand the different behavior of the time-averaged  and the
portfolio-averaged quantities, for the sake of simplicity, we consider a
portfolio composed by $N$ stocks which are  traded in a period of $T$ trading
days. We first study the properties of the two means, $\mu_i$ and  $\mu(t)$. It
is straightforward to verify that 
\begin{equation} 
<\mu_i>_i=<\mu(t)>_t\equiv\mu, 
\end{equation} 
where $<..>_t$ indicates temporal average and $<..>_i$ 
indicates ensemble average. The variances of $\mu_i$ and  $\mu(t)$ are in
general different. We obtain for the  variance of $\mu(t)$ the expression
\begin{equation}
Var[\mu(t)]_t\equiv\frac{1}{T}\sum_{t=1}^{T}(\mu(t)-\mu)^2=\frac{1}{N^2}\sum_{i=1}^N\sum_{j=1}^N
\sigma^2_{ij}, 
\end{equation}  
where $\sigma^2_{ij}$ is the return covariance between 
stock $i$ and $j$ defined as 
\begin{equation}
\sigma^2_{ij}=<R_i(t)R_j(t)>_t-<R_i(t)>_t<R_j(t)>_t. 
\end{equation} 
The width of the pdf of $\mu(t)$ (shown in  Fig.
6) is the square root of $Var[\mu(t)]_t$.
Equations (12) and (13) indicate that this quantity depends both on the 
ensemble averaged square volatility (terms with $i=j$ in Eq. (12)) and on the 
mean of the synchronous cross-covariances  between pairs of stocks
(terms with $i\ne j$ in Eq. (12)). 

With similar methods we show that
the variance of $\mu_i$ can be written as
\begin{equation}
Var[\mu_i]_i\equiv\frac{1}{N}\sum_{i=1}^{N}(\mu_i-\mu)^2
=\frac{1}{T^2}\sum_{t=1}^T\sum_{t'=1}^T \sigma^2_{tt'},
\end{equation} 
where we define the return covariance between trading day $t$ and $t'$ as
\begin{equation}
\sigma^2_{tt'}=<R_i(t)R_i(t')>_i-<R_i(t)>_i<R_i(t')>_i.
\end{equation}
This quantity gives an estimate of the correlation present in the 
whole portfolio at trading day $t$ and $t'$. 
The double sum in Eq. (14) can be split
in a term depending on the average square variety ($t=t'$) and in a term 
depending on the correlation between different trading days ($t\ne t'$).

We verify that the average square variance and volatility satisfy
the sum rule
\begin{equation}
Var[\mu_i]_i+<\sigma^2_i>_i= Var[\mu(t)]_t+<\sigma^2(t)>_t.
\end{equation}   
Combining Eq.s (12), (14) and (16) we show that
\begin{eqnarray}
\frac{T-1}{T}<\sigma^2(t)>_t+\frac{2}{N^2}\sum_{j=1}^N\sum_{i<j}
\sigma^2_{ij}= \\
=\frac{N-1}{N}<\sigma^2_i>_i+\frac{2}{T^2}\sum_{t=1}^T
\sum_{t'<t}\sigma^2_{tt'}.\nonumber
\end{eqnarray}
Since $N,T>>1$, we approximate $(N-1)/N\cong(T-1)/T\cong 1$ and Eq. (17)
becomes
\begin{equation}
<\sigma^2_i>_i-<\sigma^2(t)>_t\cong <\sigma^2_{ij}>_{i\ne j}-
<\sigma^2_{tt'}>_{t\ne t'},
\end{equation}
or equivalently
\begin{equation}
Var[\mu(t)]_t-Var[\mu_i]_i\cong <\sigma^2_{ij}>_{i\ne j}-
<\sigma^2_{tt'}>_{t\ne t'}.
\end{equation}
\begin{figure}[t]
\epsfxsize=3in
\epsfbox{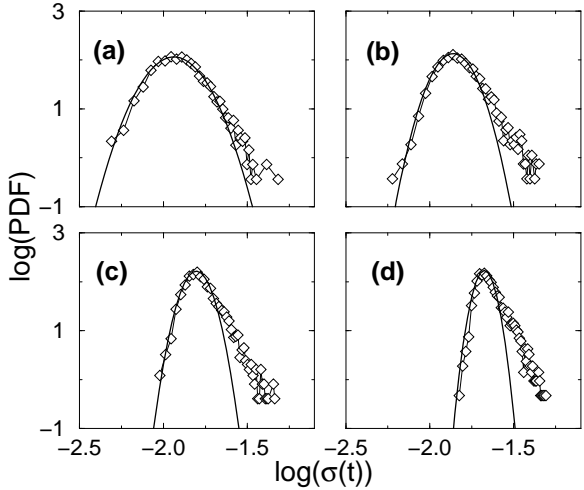}
\vspace{0.6cm}
\caption{Log-log plot of the probability density function of the variety
$\sigma(t)$ for the four considered ensemble of stocks. (a) DJIA30, (b)
SP100, (c) SP500, (d) NYSE. The solid lines are our best fit of the central
part of the distribution according to a lognormal distribution.}
\label{fig8}
 \end{figure}
Figure 6 shows that $Var[\mu(t)]_t>Var[\mu_i]_i$. 
This empirical observation together with the last relation tell us that
the synchronous cross-correlations between the stocks are on average 
stronger than the single stock correlation present in the 
whole portfolio at two different trading day. This result is consistent
with previous observations 
that synchronous returns of different stocks are significantly cross-correlated 
\cite{Elton,Campbell,Mantegna99,Bouchaud99,Stanley99}, whereas single price 
returns are poorly autocorrelated in time.
This conclusion is also verified by our empirical observation that
$<\sigma^2_i>_i><\sigma^2(t)>_t$.

\subsection{Portfolio size}     

One key aspect of the previous results concerns the degree of
generality of the observed stylized facts. In other words, are the 
empirical properties of the variety depending on the 
considered portfolio? In Section II we have shown that
all the stocks are not equivalent with respect to their
statistical properties (see the spread of points observed in Fig. 2).
In fact a trend is observed in the degree of non-Gaussian shape of the 
return distribution as a function of the stock capitalization.

To test the degree of sensitivity of our results to the average
capitalization of the selected portfolio, we repeat the  analysis
presented in subsection III.B for three other portfolios of stocks
traded in the NYSE. Specifically we investigate:  (a) the set of 30
stocks used to compute the Dow Jones Industrial Average index; (b)
the set of stocks traded in the NYSE and used to compute the
Standard \& Poor's 100 index; and (c)  the set of stocks traded in
the NYSE and   used to compute the Standard \& Poor's 500 index.
The results  obtained for all the stocks traded in the NYSE are
also  considered for reference. The four sets are different with
respect to two aspects. They differ for the number of stocks
present in the set and for the average capitalization of the
considered stocks. The empirical pdfs of $\mu(t)$ for the four considered 
sets are roughly the same. An evident different behavior is observed
for the variety. In Fig. 8 we show the pdf of the variety of the
considered portfolios of stocks. Specifically  panels (a), (b), (c)
and (d) of Fig. 8 are the results obtained for the Dow Jones 30, Standard \&
Poor's 100, Standard \& Poor's  500 and NYSE sets of stocks,
respectively. By moving from the smallest to the largest portfolio
of stocks two effects take place. The pdf of the variety becomes
progressively sharper and deviates more from a lognormal profile.
The fact that the pdf of the variety  becomes progressively sharper
is probably due to the fact the number of elements in the considered set
increases whereas we interpret the progressive deviation from the
lognormal profile as a direct manifestation of the progressive
increases of the degree of inhomogeneity of the portfolio of stocks.

In summary the presence of inhomogeneity in capitalization in  the
portfolio of stocks affects the statistical properties  of the
variety of the portfolio. This fact should be kept in  mind when
results about the variety such as results about  other statistical
properties included return distribution are obtained by considering
the statistical properties of a set of inhomogeneous stocks.

\section{Single-index model}

In this section we compare the results of our empirical analysis 
obtained for the NYSE portfolio of stocks with
the results obtained by modeling the stock price dynamics with the 
single-index model. The single-index model \cite{Elton,Campbell} is a
basic model of price dynamics in financial markets. It assumes that the
returns of all stocks are controlled by one factor,  usually called the
``market". In this model, for any stock $i$ we have  
\begin{equation}
R_i(t)=\alpha_i+\beta_i R_M(t)+\epsilon_i(t),  
\end{equation}  
where $R_i(t)$ and $R_M(t)$ are the return of the stock $i$ and of the 
``market" at day $t$, respectively, $\alpha_i$ and $\beta_i$ are two
real  parameters and $\epsilon_i(t)$ is a zero mean noise term
characterized by a variance equal to $\sigma^2_{\epsilon_i}$. The noise
terms of different  stocks are assumed to be uncorrelated,
$<\epsilon_i(t) \epsilon_j(t)>_t=0$  for $i\neq j$.  Moreover the
covariance between $R_M(t)$ and $\epsilon_i(t)$ is set to zero for any
$i$. 

Each stock is correlated with the market and the presence of such
a  correlation induces a correlation between any pair of stocks. 
It is customary to adopt a broad-based stock index for the market
$R_M(t)$.  Our choice for the ``market" time series is the Standard
and Poor's 500 index. The best estimation of the model parameters
$\alpha_i$, $\beta_i$ and $\sigma^2_{\epsilon_i}$ is done with the
ordinary least squares method \cite{Campbell}. 
In order to compare our empirical
results with those 
\begin{figure}[t]
\epsfxsize=3in
\epsfbox{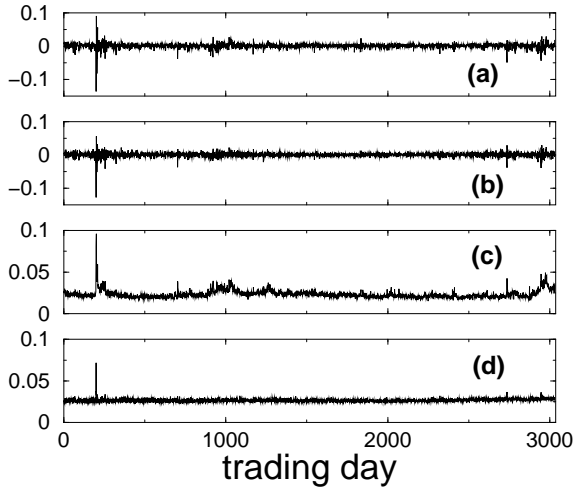}
\vspace{0.6cm}
\caption{(a) Time series of the mean of the ensemble return
distribution $\mu(t)$. (b) Time series of the mean of the ensemble
return distribution  for the surrogate data generated according to
the single-index model. (c) Time series of the variety $\sigma(t)$
of the ensemble return distribution. (d) Time series of the variety
of the ensemble return distribution for the surrogate data
generated according to the single-index model.}
\label{fig9}
 \end{figure}
predicted by the single-index model  we build up
an artificial market according to Eq. (20). To this end we first
evaluate the model parameters for all the stocks traded in the
NYSE  and then we generate a set of $n$ of surrogate time series
according to  Eq. (20). To make the simulation as realistic as
possible, in the  generation of our surrogate data set we use as
``market"  time series the true time series of the  Standard and
Poor's 500 index.

We evaluate the central moments $\mu(t)$ and $\sigma(t)$ defined in
Eqs (7-8) for the surrogate data. In Fig. 9(a) we show the time
series of $\mu(t)$  of the real data and in Fig. 9(b) we show the
same quantity for the surrogate market data generated according to
the single-index model. The agreement between the two time series
is pretty  high and therefore the single-index model describes quite
well the mean returns of the market at time $t$ provided that the
behavior of the ``market" $R_M(t)$ is known . This result is also
confirmed  by Fig. 10 where the pdf of $\mu(t)$ for real and
surrogate data are shown. Also the time correlation properties of 
surrogate $\mu(t)$ are pretty similar to the real ones.
In fact, a fast decaying autocorrelation function of $\mu(t)$ is 
observed in surrogate data. A good agreement is also observed 
when one investigates 
the statistical properties of $\mu_i$ and $\sigma_i$. The 
single-index model approximates quite well the empirical 
distribution of $\mu_i$ and $\sigma_i$.

A different behavior is observed for the variety $\sigma(t)$.
Figure 9(c) and 9(d) show the time series of $\sigma(t)$ for real
and surrogate data, respectively. The real time series of the
variety is non stationary and shows several bursts of activity. On
the contrary the surrogate time series  is quite stationary with
the exception of the 1987 crash. Figure 11 shows  the pdfs of
$\sigma(t)$   for real and surrogate data. The model fails in
describing the distribution of $\sigma(t)$.
\begin{figure}[t]
\epsfxsize=3in
\epsfbox{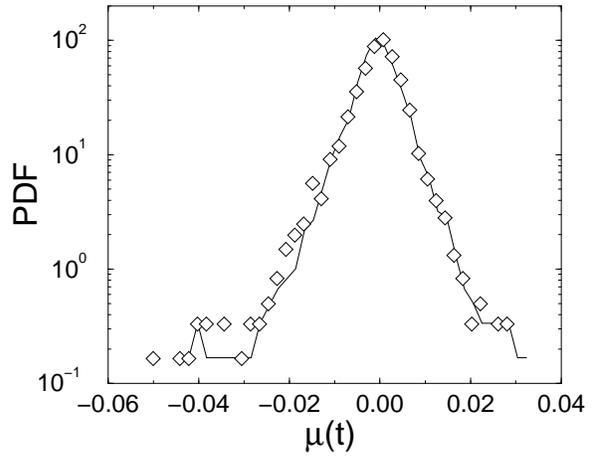}
\vspace{0.5cm}
\caption{Comparison of the probability density function
of the mean $\mu(t)$ of the ensemble return distribution 
obtained from real (diamond) with the one obtained from surrogate 
data generated according to the single-index model(continuous line).}
\label{fig10}
 \end{figure}
\begin{figure}[t]
\epsfxsize=3in
\epsfbox{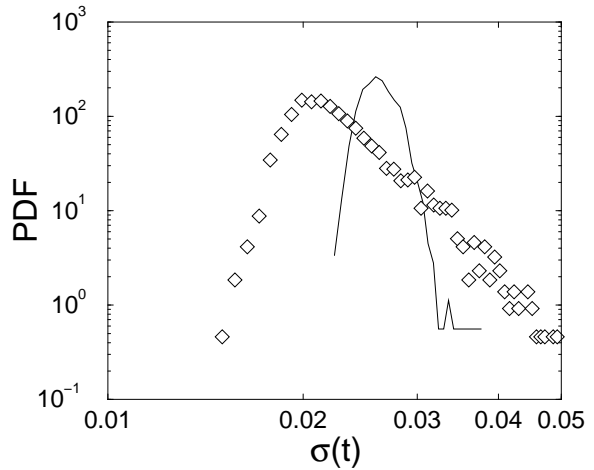}
\vspace{0.5cm}
\caption{Comparison of the probability density function of the
variety $\sigma(t)$ obtained from real (diamond) with the one obtained from
surrogate data generated according to the single-index model 
(continuous line).}
\label{fig11}
 \end{figure}

In summary, the single-index model gives a good  approximation of
the statistical behavior of $\mu(t)$, $\mu_i$ and $\sigma_i$
whereas it describes poorly the  statistical behavior of the
variety of a portfolio of stocks traded in a financial market. This
conclusion is also supported by the observation that  the
autocorrelation function of the surrogate variety decays  in $2-3$
trading days to the value $0.1$ and the power spectrum is very
similar to a white noise spectrum, whereas long-range correlation
is  observed in real data.  

A more refined analysis shows that  the artificial
ensemble return distribution is systematically less leptokurtic
than the real one.  Moreover, in Ref. \cite{epjb00} we show that the 
single-index model is unable to predict the change in the symmetry
properties  of the ensemble return distribution in crash and rally
days. The differences observed between the
behavior of real data and the behavior of surrogate data suggest
that the correlations among the stocks can be explained by the
single-index model only for ``normal" periods 
in first approximation  whereas the model miss completely to reproduce the
correlation behavior during extreme events.

\section{Conclusions}

The present study shows that one needs to consider not only  the
statistical properties characterizing the time evolution of price
for each stock traded but also the synchronous  collective behavior
of the portfolio considered to reveal the  overall complexity of a
financial market. We show that  such a collective behavior of a
portfolio of stock is  efficiently monitored by the variety of the
ensemble  return distribution.  This variable is directly
observable for each portfolio and presents interesting statistical
properties. It is non-Gaussian distributed and long-range
correlated. The detailed statistical properties  depends on the
considered portfolio of stocks. We verify that  for a portfolio of
stocks characterized by comparable capitalization  the distribution
of the variety is approximately lognormal. Deviation from the
lognormal behavior are observed for less homogeneous  (in
capitalization) portfolios.

The shape of the distribution and the long-term memory of the variety are
not reproduced by considering surrogated data simulated by using a 
single-index model with a realistic time series for the ``market". This implies
that the complexity detected by the performed empirical analysis
cannot  be modeled with a similar simple stock price model. The
correlations present in the market are more complex than the ones
hypothesized  by the single-index model. 

The correct modeling of the statistical properties of the variety
can be then used as a benchmark for stock price models more 
sophisticated than the single-index model. 

The ensemble return distribution shows a qualitatively and
quantitatively different behavior in ``normal" and extreme trading
days. The variety of a portfolio is then able to detect quite
clearly shocks and  aftershocks occurring in the market. Hence, it
is a promising direct observable able to measure how much a portfolio is
under pressure and how distant is from the typical market activity in a 
specific trading day. A theoretical challenge is to relate this empirical 
ensemble observation directly with the correlations active between pairs of
stocks of a correlation.

In summary, we believe that the overall complexity of a financial
market can be detected and modeled only by considering
simultaneously -- (i) the statistical properties of the time evolution of
stock prices of the considered portfolio {\it and} (ii) the statics
and dynamics of the correlations existing between stocks.

\section{Acknowledgements}
The authors thank INFM and MURST for financial support. This work 
is part of the FRA-INFM project {\it Volatility in financial markets}. 
F. Lillo acknowledges FSE-INFM for his fellowships.
We wish to thank Giovanni Bonanno for help in numerical calculations.

%----------------------------------------------------------------------
%\newpage
%\centerline{\bf References}

%----------------------------------------------------------------------  
\end{document}